\documentclass[letterpaper,12pt]{article}
\usepackage{tabularx} 
\usepackage{amsmath,bm}  
\usepackage{float}
\usepackage{indentfirst}
\usepackage{graphicx} 
\usepackage[margin=1in,letterpaper]{geometry} 
\usepackage{caption}
\usepackage{cite} 
\usepackage[colorlinks,linkcolor=magenta,anchorcolor=cyan,citecolor=blue]{hyperref} 
\usepackage{tensor}
\hypersetup{
	colorlinks=true,       
	linkcolor=blue,        
	citecolor=blue,        
	filecolor=magenta,     
	urlcolor=blue
}

\def\be{\begin{equation}}
\def\ee{\end{equation}}
\def\ba{\begin{eqnarray}}
\def\ea{\end{eqnarray}}

              \def\.{\cdot}

\begin{document}
\begin{center}

	\vspace{10pt}
	\large{\bf{The linearized second law for any higher curvature gravity with the scalar and the electromagnetic fields}}
	
	\vspace{15pt}
	Xin-Yang Wang $^\text{a}$, Jie Jiang $^\text{a}$
	
	\vspace{15pt}
	\small{\it $^\text{a}$ College of Education for the Future, Beijing Normal University, Zhuhai 519087, China}
	\vspace{30pt}
\end{center}
\begin{abstract}
	The first law of black hole thermodynamics is suitable for any diffeomorphism invariant gravity, and the entropy in the first law is the Wald entropy which is highly dependent on the non-minimal coupling interactions in the theory of gravity. However, whether the Wald entropy still satisfies the second law needs to be investigated. The entropy of black holes obeying the linearized second law in arbitrary high-order curvature gravity is given, which can be written as the Wald entropy with correction terms. It indicates that the Wald entropy is not commonly obeying the linearized second law for any high-order curvature gravity. When the interactions of gravity with matter fields are included in the theory of gravity, the entropy of black holes obeying the linearized second law has not been obtained in this case. Considering any high-order curvature gravity with the scalar and the electromagnetic fields, from the Raychaudhuri equation, the entropy obeying the linearized second law is generally obtained, which can be expressed as the Wald entropy with correction terms as well. The entropy does not include the contribution from the electromagnetic fields, and the correction terms contain the contribution from the minimal coupling interaction between gravity and the scalar fields. Since the entropy satisfying the linearized second law depends only on the non-minimal coupling interaction of gravity in previous research, this result upends our understanding of the entropy of black holes obeying the linearized second law in any gravitational theory with matter fields.
\end{abstract}
	\vfill {\footnotesize ~\\ $^\text{a}$ xinyangwang@bnu.edu.cn \\ Corresponding author: \\ $^\text{a}$ jiejiang@mail.bnu.edu.cn}
\newpage

\section{Introduction}
Black holes as a particular spacetime structure have been predicted by general relativity. For black holes in classical general relativity, the singularity lies at the center of black holes and is surrounded by a specific null hypersurface. The null hypersurface is called the event horizon of black holes and is usually regarded as the boundary of black holes. The event horizon plays a critical role in the investigation of black hole physics because many essential properties of black holes, especially black hole thermodynamics, are reflected by the event horizon in an equilibrium state or a state of dynamic evolution. The area law of black holes was suggested first by Hawking \cite{Hawking:1971tu}, which states that the area of the event horizon of black holes never decreases along the direction of time evolution. From the perspective of the area law, Bekenstein \cite{Bekenstein:1973ur} proposed that the area of the event horizon could be identified with the entropy of the classical adiabatic thermodynamic system directly because the evolution tendency of the area of the event horizon is similar to the evolution of the entropy constrained by the second law of thermodynamics. From the quantum field theory in curved spacetime, Hawking \cite{Hawking:1974sw} first proved that the entropy and the temperature of black holes are defined respectively by the area of the event horizon and the surface gravity of black holes. And the entropy and the area of the event horizon of black holes satisfy a simple proportional relationship, i.e., $S_{\text{BH}} = A / 4$, where $A$ is the area of the event horizon of black holes. The entropy satisfying the proportional relationship is called Bekenstein-Hawking entropy. According to the definitions of the temperature and the entropy of black holes, the four laws of black hole thermodynamics are established \cite{Bekenstein:1972tm, Bardeen:1973gs, Unruh:1976db}. Since the four laws of black hole thermodynamics are identical to the four laws of thermodynamics satisfied by the classical thermodynamic system, black holes can be regarded as thermodynamic systems rather than pure spacetime structures. In the four laws of black hole thermodynamics, the two profound laws for black holes are respectively the first and second laws. If black holes are regarded as thermodynamic systems, the two laws of thermodynamics should be seen first as robust features of black holes. Although black holes in classical general relativity automatically satisfy the two laws of thermodynamics, the two laws of thermodynamics are not always guaranteed to hold for black holes in any diffeomorphism invariant gravitational theory. An issue of whether the first and the second laws of thermodynamics are still the best features of black holes in any gravitational theory is raised naturally. Starting with this issue, Iyer and Wald \cite{Wald:1993nt, Iyer:1994ys} proposed the variation method first to investigate black hole thermodynamics in arbitrary diffeomorphism invariant gravitational theory. The result shows that the first law of thermodynamics is generally suitable for black holes. And the entropy that always matches the first law of black hole thermodynamics is called the Wald entropy rather than Bekenstein-Hawking entropy. Although it has been demonstrated that the Wald entropy is generally suitable for the first law of black hole thermodynamics in any gravitational theory, whether it always satisfies the second law of thermodynamics has not been investigated enough. Therefore, in the following, we will discuss mainly the matching relationship between Wald entropy and the second law of thermodynamics.

So far, a suitable scheme to quantize the gravitational theory has not been established. It means that if the self-interactions of gravity or the interactions of gravity with matter fields are included in the gravitational theory, the two kinds of interactions are not studied rigorously at the complete quantum regime. To investigate the interactions in the frame of quantum gravity, one of the most fruitful methods is to construct the corresponding low-energy efficient gravitational theory to study the two types of interactions approximately. This method essentially introduces high-order curvature terms and minimal or non-minimal coupling interaction terms between gravity and matter fields, which correspond to two kinds of interactions and are called the quantum correction terms of gravity, into the effective Lagrangian of the quantum gravity theory at a low energy scale. In other words, when the low-energy efficient scheme is adopted to deal with the issue of the quantization of gravity, some quantum correction terms, which are the high-order curvature terms and the minimal or non-minimal coupling interaction terms between gravity and matter fields, should be added to the Lagrangian of the gravitational theory \cite{Capozziello:2011et}.

The expression of the Wald entropy is closely dependent on the Lagrangian of the gravitational theory according to the definition. When the quantum correction terms that correspond to the two types of interactions are included in the Lagrangian, the non-minimal coupling interaction terms in these quantum correction terms will sufficiently influence the expression of the Wald entropy. We can infer that a substantial change in the specific expression of the Wald entropy will inevitably affect the matching relationship between the Wald entropy and the second law of thermodynamics. Therefore, to investigate whether the Wald entropy generally satisfies the second law, one should consider the effect of each term in the Lagrangian of the gravitational theory describing non-minimal coupled interactions on the matching relationship between the Wald entropy and the second law before finding a general research method. Fortunately, for the case that only high-order curvature terms which describe the self-interactions of gravity exist in the Lagrangian, a research technique that can generally investigate the linearized second law under the matter fields perturbation is proposed by A. C. Wall \cite{Wall:2015raa}. The results show that the Wald entropy does not always satisfy the linearized second law in any high-order curvature gravity, and the general form of the entropy of black holes obeying the linearized second law is given, which can be expressed as the Wald entropy with correction terms. However, for the other case that the non-minimal coupling interactions between gravity and matter fields are contained in the Lagrangian of the gravitational theory, the second law of black hole thermodynamics has not been investigated enough, and the general expression of the entropy of black holes that satisfies the linearized second law has not been obtained yet. In our previous works on the linearized second law, we have mainly considered the gravitational theory with the non-minimal coupling interaction terms between gravity and matter fields. For the Horndeski gravity and the general quadric corrected Einstein-Maxwell gravity, we have shown that the Wald entropy obeys the linearized second law of thermodynamics during the matter field perturbation \cite{Wang:2020svl, Wang:2021zyt}. However, when investigating the linearized second law in general second-order scalar-tensor gravity, we found that the evolution of Wald entropy during the perturbation process no longer satisfies the requirements of the linearized second law. And the expression of entropy obeying the linearized second law can also be written as the form of the Wald entropy with relevant correction terms \cite{Wang:2022stg}. According to this result, we can infer that the Wald entropy does not always obey the linearized second law for any gravitational theory with the non-minimal coupling interactions between gravity and matter fields. Therefore, to study the linearized second law of black hole thermodynamics in any diffeomorphism invariant gravity with the interactions between gravity and matter fields, the general expression of the entropy of black holes satisfying the linearized second law should be obtained first.

The scalar fields in gravitational theory have been a topic of great interest in recent years because scalar field dynamics can help us understand some detailed features of the universe. From an empirical motivation mainly related to astronomical observations, the scalar fields can be regarded as a powerful tool to explain many phenomena at the galactic and cosmological scales. It means that the gravitational theories incorporating scalar fields may help us understand these phenomena, such as the origin of the early universe and its late-time accelerated expansion, as well as the presence of dark matter and dark energy \cite{Cisterna:2014nua}. Meanwhile, the properties of these phenomena also confirm that the scalar fields are suitable candidates to solve these unknown phenomena in the universe. Therefore, many gravitational theories containing scalar fields, such as Brans-Dicke theory \cite{Brans:1961sx}, inflation theory, and several other cosmological models \cite{Bertacca:2007ux, Chung:2007vz, delaMacorra:2007beq, Saridakis:2010mf}, are gradually established. Besides, from the perspective of astronomical observation, many celestial bodies in our universe commonly take electric charge, and outside spacetime fills with the electromagnetic fields rather than the vacuum. It means that the electromagnetic fields should be contained generally in the theory of gravity. Therefore, based on the above facts, and according to the gravitational theories used to study the linearized second law in our previous works, we will consider the linearized second law of black hole thermodynamics in a general diffeomorphism invariant high-order curvature gravitational theory containing the minimal and non-minimal interactions of gravity with the scalar and the electromagnetic fields. Furthermore, we will derive the expression of the entropy of black holes obeying the linearized second law in this general gravitational theory during the matter field perturbation.

The organization of the paper is as follows. In Sec. \ref{sec2}, a general diffeomorphism invariant high-order gravitational theory with the scalar and the electromagnetic fields is introduced. Furthermore, the definition of the Wald entropy and its specific expression in the gravitational theory will be given, respectively. In Sec. \ref{sec3}, a perturbation process is introduced first, which comes from the additional matter fields outside black holes, to investigate the linearized second law. Based on the assumptions that the matter fields should satisfy the null energy condition and that a regular bifurcation surface exists in the background spacetime, we will derive the general expression of the entropy of black holes obeying the linearized second law of thermodynamics in the general diffeomorphism invariant gravity during the perturbation process. The paper ends with discussions and conclusions in Sec. \ref{sec4}.

\section{The general high-order curvature gravity with the scalar and the electromagnetic fields and the Wald entropy}\label{sec2}
 
We consider an arbitrary diffeomorphism invariant high-order curvature gravitational theory with the scalar and the electromagnetic fields to investigate the linearized second law of black hole thermodynamics, and the general expression of the entropy of black holes satisfying the linearized second law in the gravitational theory will be further derived. The Lagrangian of the general diffeomorphism invariant gravity can be expressed formally as
\begin{equation}
	\mathcal{L} = \mathcal{L} \left(g_{ab}, R_{abcd}, F_{ab}, \phi, \nabla_a \phi, \nabla_a \nabla_b \phi \right)\,.
\end{equation}
To obtain the general expression of the entropy of black holes obeying the linearized second law in this gravitational theory, the additional matter fields that are minimally coupling with gravity should be introduced. Moreover, a quasistationary process is further involved, which states that the matter fields existing outside black holes pass through the event horizon and fall into the interior of black holes. The spacetime configuration of black holes can be perturbed through the matter fields during the process. The point of view indicates that the matter fields and the spacetime of black holes can be regarded as a complete dynamical system, and the perturbation process is a dynamic evolution process of the system. Therefore, when the additional matter fields appear in spacetime, the Lagrangian of the gravitational theory can be expanded as
\begin{equation}\label{usedlagrangian}
	\mathcal{L} = \mathcal{L} \left(g_{ab}, R_{abcd}, F_{ab}, \phi, \nabla_a \phi, \nabla_a \nabla_b \phi \right) + \mathcal{L}_{\text{mt}}\,,
\end{equation}
where $\mathcal{L}_{\text{mt}}$ represents the Lagrangian of the additional matter fields in spacetime. After calculating the variation of the expanded Lagrangian with respect to the metric $g_{ab}$, the equation of motion of the gravitational part can be formally expressed as 
\begin{equation}\label{equationofmotion}
	H_{ab} = 8 \pi T_{ab}\,.
\end{equation}
The left-hand side of Eq. (\ref{equationofmotion}) can be further written as a linear combination of four components,
\begin{equation}\label{fourpartsofeom}
	H_{ab} = H_{ab}^{1} + H_{ab}^{2} + H_{ab}^{3} + H_{ab}^{4} \,.
\end{equation}
The first part corresponds to the derivative of the Lagrangian to the Riemann curvature $R_{abcd}$. The second part comes from the derivative of the Lagrangian to the first-order covariant derivative of the scalar field $\nabla_a \phi$. The third part is derived from the derivation of the Lagrangian to the second-order covariant derivative of the scalar field $\nabla_a \nabla_b \phi$. The fourth part is the derivative of the Lagrangian to the electromagnetic field $F_{ab}$. The specific expressions of the four components in Eq. (\ref{fourpartsofeom}) are expressed as
\begin{equation}\label{specexprefourpartofeom}
	\begin{split}
		H_{ab}^{1} & = (E_R)_{a}^{\ cde} R_{bcde} + 2 \nabla^c \nabla^d \left(E_R \right)_{acbd}\,, \\
		H_{ab}^{2} & = \frac{1}{2} \left(E_1 \right)_a \nabla_b \phi\,, \\
		H_{ab}^{3} & = - \nabla^c \left(E_2 \right)_{cb} \nabla_a \phi + \frac{1}{2} \nabla^c \left(E_2 \right)_{ab} \nabla_c \phi + \frac{1}{2} \left(E_2 \right)_{ab} \nabla_c \nabla^c \phi\,, \\
		H_{ab}^{4} & = \left(E_F\right)_{a}^{\ c} F_{bc}\,,
	\end{split}
\end{equation}
where
\begin{equation}
	\left(E_R \right)^{abcd} = \frac{\partial \mathcal{L}}{\partial R_{abcd}}\,, \quad \left(E_1 \right)^a = \frac{\partial \mathcal{L}}{\partial \nabla_a \phi}\,, \quad \left(E_2 \right)^{ab} = \frac{\partial \mathcal{L}}{\partial \nabla_a \nabla_b \phi}\,, \quad \left(E_F \right)^{ab} = \frac{\partial \mathcal{L}}{\partial F_{ab}}\,.
\end{equation}
Besides, on the right-hand side of Eq. (\ref{equationofmotion}), $T_{ab}$ is the stress-energy tensor of the theory of gravity, which only contains the stress-energy tensor of the additional matter fields $T_{ab}^{\text{mt}}$. From the physical perspective, we assume that the additional matter fields should satisfy the null energy condition during the perturbation process. Based on the assumption and the fact that the total stress-energy tenor of the theory of gravity only contains the stress-energy tensor of the additional matter fields, for any null vector field $n^a$ in spacetime, the null energy condition can be expressed as
\begin{equation}
	T_{ab}^{\text{mt}} n^a n^b = T_{ab} n^a n^b \ge 0\,.
\end{equation}

In a $(n+2)$-dimensional general diffeomorphism invariant gravitational theory described by the Lagrangian in Eq. (\ref{usedlagrangian}), the Wald entropy of black holes in stationary background spacetime of the gravitational theory can be defined as
\begin{equation}\label{definewaldentropy}
	S_{\text{W}} = \frac{1}{4} \int_s d^{n} y \sqrt{\gamma} \rho_{\text{W}}\,,
\end{equation}
where $\gamma$ is the determinant of the induced metric on any slice of the event horizon, $y$ is introduced to label the transverse coordinates on the cross-section, and $\rho_{\text{W}}$ is the entropy density of the Wald entropy. While the entropy density is further given as
\begin{equation}
	\rho_{\text{W}} = - 8 \pi \frac{\partial \mathcal{L}}{\partial R_{abcd}} \epsilon_{ab} \epsilon_{cd}\,.
\end{equation}
In which $\mathcal{L}$ is the Lagrangian of the gravitational theory, $R_{abcd}$ is the tensor of the Riemann curvature, and $\epsilon_{ab}$ is the binormal on any cross-section of the event horizon. From the definition of the Wald entropy, one can see that when the scalar and the electromagnetic fields are both contained in the gravitational theory, the matter fields that only participate in the non-minimal coupling interaction with gravity can sufficiently affect the expression of the Wald entropy. Meanwhile, the matter fields that involve the minimal coupling interaction with gravity do not influence the specific expression of the Wald entropy. For an arbitrary high-order curvature gravitational theory, it has been shown that the Wald entropy of black holes in the gravitational theory does not always obey the linearized second law during the matter field perturbation. While the entropy of black holes always satisfies the linearized second law in the gravitational theory containing only high-order curvature terms have been generally given, which can be expressed as the Wald entropy with correction terms. Therefore, in the following, considering the high-order curvature gravitational theory with the scalar and the electromagnetic fields, starting with the definition of the Wald entropy again, we should find out how to modify the expression of the Wald entropy such that it always satisfies the linearized second law under the first-order approximation of the matter field perturbation. And finally, we would like to obtain the general expression of the entropy of black holes obeying the linearized second law in any diffeomorphism invariant high-order curvature gravitational theory with the scalar and the electromagnetic fields.

\section{The linearized second law for any high-order curvature gravity with the scalar and the electromagnetic fields}\label{sec3}

As mentioned above, a physical quasistationary accretion process of black holes is introduced to investigate the linearized second law of black hole thermodynamics. The accretion process describes the dynamical process where the additional matter fields minimally coupling to gravity fall into black holes and perturb the spacetime geometry of black holds. To obtain the general expression of the entropy of black holes satisfying the linearized second law in any diffeomorphism invariant high-order curvature gravitational theory with the scalar and the electromagnetic fields during the matter fields perturbation, we should assume that black holes will finally settle down to a stationary state after the matter field perturbation process. This assumption is called the stability assumption for simplicity in the following.

For the $\left(n +2 \right)$-dimensional diffeomorphism invariant high-order curvature gravitational theory with the scalar field, the electromagnetic field, and the additional matter fields described by the Lagrangian in Eq. (\ref{usedlagrangian}), the event horizon of black holes which is $\left(n +1 \right)$-dimensional hypersurface in spacetime is denoted as $\mathcal{H}$. A parameter ``$u$'' is introduced as an affine parameter to parameterize the event horizon. Furthermore, a null vector field $k^a = \left(\partial / \partial u \right)^a$ can be chosen to generate the event horizon and satisfies the geodesic equation $k^b \nabla_b k^a = 0$. Any specific value of the parameter ``$u$'' corresponds to a cross-section on the event horizon. We can establish coordinates with two null vectors, i.e., $\{k^a, l^a, y^a \}$, on the cross-section of the event horizon. In the coordinates, the null vector $l^a$ is another null vector which is different from $k^a$, and another parameter ``$v$'' is chosen to represent the null vector $l^a$ as $l^a = \left(\partial / \partial v \right)^a$. These null vectors in the coordinates satisfy the following two relationships,
\begin{equation}
	k^a k_a = l^a l_a = 0, \qquad k^a l_a = - 1\,.
\end{equation}
According to the two null vector fields, the binormal on any cross-section is defined as $\epsilon_{ab} = 2 k_{[a} l_{b]}$, and the induced metric on any slice of the event horizon is defined as  
\begin{equation}\label{inducedmetric}
	\gamma_{ab} = g_{ab} + 2 k_{(a} l_{b)}\,.
\end{equation}
From the induced metric and the null vector fields $k^a$, the extrinsic curvature of the event horizon can be defined by 
\begin{equation}\label{extrinsiccurva}
	B_{ab} = \gamma_{a}^{\ c} \gamma_{b}^{\ d} \nabla_c k_d\,.
\end{equation}
Using the definition of the induced metric and the expression of  the extrinsic curvature in Eq. (\ref{inducedmetric}) and Eq. (\ref{extrinsiccurva}), the evolution of the induced metric along the direction of the future event horizon can be given as
\begin{equation}
	\gamma_{a}^{\ c} \gamma_{b}^{\ d} \mathcal{L}_k \gamma_{cd} = 2 \left(\sigma_{ab} + \frac{\theta}{n} \gamma_{ab} \right) = 2 B_{ab}\,,
\end{equation}
where $\sigma_{ab}$ and $\theta$ represent the shear and the expansion of the event horizon during the evolutionary process, respectively. Moreover, the evolution of the extrinsic curvature is obtained as 
\begin{equation}\label{evoluextriccurve}
	\gamma_{a}^{\ c} \gamma_{b}^{\ d} \mathcal{L}_k B_{cd} = B_{ac} B_{b}^{\ c} - \gamma_{a}^{\ c} \gamma_{b}^{\ d} R_{ecfd} k^e k^f\,.
\end{equation}
Utilizing the evolution property of the extrinsic curvature, the Raychaudhuri equation can be further given as 
\begin{equation}\label{rayequation}
	\frac{d \theta}{d \lambda} = - \frac{\theta^2}{n - 2} -\sigma_{ab} \sigma^{ab} - R_{uu}\,,
\end{equation}
where the quantity $R_{uu}$ is the abbreviation of $k^a k^b R_{ab}$. In the following, we will use some Latin letters at the beginning of the alphabet, i.e., $a$, $b$, $c$, $\cdots$, to represent the abstract index in any tensor and will use some Latin letters which start from the letter $i$, i.e., $i$, $j$, $k$, $\cdots$, to represent the spatial index in any tensor. Meanwhile, a convention will be further introduced to simplify the expressions of equations. This convention can be stated as follows.
\\
\\
\noindent (1) An index in any tensor can be replaced by the parameter $u$ or $v$ directly when the index contracts with the null vector $k^a$ or $l^a$.
\\
\\
\noindent (2) An index in any tensor can be replaced by one of the spatial indexes, i.e., $i$, $j$, $k$, $\cdots$, when the index contracts with the induced metric on the cross-section of the event horizon.
\\ 
\\
\noindent According to the convention, for any tensor $X_{a_1 b_1 \cdots a_2 b_2 \cdots a_3 b_3 \cdots}$, when indexes, $a_1$, $b_1$ $\cdots$, in the tensor contract with the null vector field $k^a$, indexes, $a_2$, $b_2$, $\cdots$, in the tensor contract with the null vector field $l^a$, and indexes, $a_3$, $b_3$, $\cdots$, in the tensor contract with the induced metric on the cross-section of the event horizon, the contracted tensor can be simplified as 
\begin{equation}
	k^{a_1} k^{b_1} \cdots l^{a_2} l^{b_2} \cdots \gamma_{c_3(i)}^{\ a_3} \gamma_{d_3(j)}^{\ b_3} \cdots X_{a_1 b_1 \cdots a_2 b_2 \cdots a_3 b_3 \cdots} = X_{u u \cdots v v \cdots i j \cdots}\,.
\end{equation}

To explicitly depict the perturbation process caused by the additional matter fields outside black holes, a sufficient small parameter $\epsilon$ will be introduced to represent the order of approximation of the perturbation. We can assume that the three quantities, which contain the extrinsic curvature, the expansion, and the shear of the event horizon, contribute only under the first-order approximation of the matter field perturbation. Utilizing the small parameter $\epsilon$, the relationship of the three quantities under the first-order approximation can be written as $B_{ab} \sim \theta \sim \sigma_{ab} \sim \mathcal{O} \left(\epsilon \right)$. Since we hope to find out the general expression of the entropy of black holes obeying the linearized second law in any high-order curvature gravitational theory with the scalar and the electromagnetic fields, the symbol ``$\simeq$'' will be used to represent the identity under the first-order approximation of the perturbation process. According to the above conventions, the extrinsic curvature of the event horizon and the evolution of the extrinsic curvature along the future event horizon, which have been given in Eq. (\ref{extrinsiccurva}) and Eq. (\ref{evoluextriccurve}), can be rewritten as 
\begin{equation}\label{bandevlob}
	B_{ij} \simeq D_{i} k_{j}, \quad \mathcal{L}_k B_{ij} \simeq  - R_{uiuj}\,,
\end{equation}
under the linear order approximation, where the derivative operator $D_a$ is the pure spatial derivative operator. For any tensor $X_{a_1 a_2 \cdots}$, the spatial derivative operator can be defined as 
\begin{equation}
	D_a X_{a_1 a_2 \cdots} = \gamma_{a}^{\ b} \gamma_{a_1}^{\ b_1} \gamma_{a_2}^{\ b_2} \cdots \nabla_{b} X_{b_1 b_2 \cdots}\,.
\end{equation}
While the linearized version of the Raychaudhuri equation can be further written as 
\begin{equation}
	\frac{d \theta}{d \lambda} \simeq - R_{uu}\,.
\end{equation}
According to the induced metric and the null vector fields $l^a$, a new quantity $C_{ij}$ can be defined as $C_{ij} = D_{i} l_{j}$. Following the calculation method of the evolution of the extrinsic curvature along the future direction of the event horizon, the evolution of $C_{ij}$ along the same direction on the background spacetime can be calculated as
\begin{equation}\label{evolutionc}
	\mathcal{L}_k C_{ij} = - R_{iujv}\,.
\end{equation}
Besides, according to the definition of the Wald entropy and the above conventions, the entropy density of the Wald entropy $\rho_{\text{W}}$ in coordinates with two null vectors $k^a$ and $l^a$ can be given as
\begin{equation}\label{entropydensity}
	\rho_{\text{W}} = - 2 \left(E_R \right)_{uvuv}\,.
\end{equation}
The null energy condition of the total stress-energy tensor of the gravitational theory in the same coordinates can be rewritten as
\begin{equation}\label{nullenergycondkk}
	T_{ab} k^a k^b = T_{uu} \geq 0\,.
\end{equation}

Next, we will investigate the linearized second law of black hole thermodynamics in the general diffeomorphism invariant high-order curvature gravitational theory with the scalar and the electromagnetic fields. Starting with the definition of the Wald entropy, the expression of the entropy of black holes that always satisfies the linearized second law in the theory of gravity under the first-order approximation of the matter field perturbation should be further derived. If the linearized second law in this gravitational theory holds, in other words, the value of the general expression of the entropy monotonously increases with the matter fields perturbation process, the entropy of black holes should satisfy the following relationship under the linear order approximation of the perturbation process \cite{Kolekar:2012tq}, i.e.,
\begin{equation}\label{relationshipentropy}
	\mathcal{L}_k^2 S \simeq - \frac{1}{4} \int_s \tilde{\epsilon} H_{uu} = - 2 \pi \int_s \tilde{\epsilon} T_{uu} \le 0\,,
\end{equation}
where $S$ represents the entropy of black holes, and the equation of motion of the gravitational part and the null energy condition of the stress-energy tensor in Eq. (\ref{equationofmotion}) and Eq. (\ref{nullenergycondkk}) have been used in the second and the third steps. The stability assumption requires that the rate of change of the entropy gradually decreases to zero after the perturbation process. The variation tendency of the rate of change of the entropy can be expressed equivalently as the second-order Lie derivative of the entropy is always negative during the perturbation process, i.e., $\mathcal{L}_k^2 S \le 0$. Combining the negative second Lie derivative of the entropy with the stability assumption, one can see that the first-order Lie derivative of the entropy should always be positive during the perturbation process, $\mathcal{L}_k S \ge 0$. It means that the value of the entropy is monotonously increasing with the matter field perturbation process. Therefore, if the entropy of black holes obeys the relationship in Eq. (\ref{relationshipentropy}), the entropy will always satisfy the linearized second law of black hole thermodynamics under the perturbation process. To obtain the general expression of the entropy of black holes that obeys the negative second-order Lie derivative in the general high-order curvature gravitational theory with the scalar and the electromagnetic fields, according to the relationship, we should calculate the specific expression of $H_{uu}$ under the linear order approximation firstly.

As mentioned above, two assumptions have been suggested. The first is that the total stress-energy tenor should obey the null energy condition, and the second is the stability assumption. However, before calculating the expression of $H_{uu}$ under the linear order approximation of the perturbation process, an additional assumption should be introduced, which states that a regular bifurcation surface exists in the background spacetime. The regular property means that all physical quantities are smooth and finite on the whole Killing horizon, even on the bifurcation surface. For the coordinates with two null vectors $k^a$ and $l^a$, $\{k^a, l^a, y^a \}$, an arbitrary vector $z_{i}^{a}$ can be introduced, which represents one of the two null vectors in the coordinates and can be expressed as $z_{i}^{a} \in \{k^a, l^a\}$. For any tensor $X_{a_1 \cdots a_k}$ which is smooth and finite on the whole Killing horizon with the bifurcation surface, the assumption of the regular bifurcation surface implies that after contracting all indexes in the tensor $X_{a_1 \cdots a_k}$ with the vectors $z_{i}^{a}$, i.e., $X_{a_1 \cdots a_k} z_{1}^{a_1} \cdots z_{k}^{a_k}$, this quantity will vanish on the background spacetime if the number of $k^a$ including in the vector $z_{i}^{a_i}$ is larger than the number of $l^a$ including in the vector $z_{i}^{a_i}$. Hence, the quantity on the background spacetime can be written as
\begin{equation}
	X_{a_1 \cdots a_k} z_{1}^{a_1} \cdots z_{k}^{a_k} = 0\,.
\end{equation}
Besides, the quantity is still finite and smooth over the whole Killing horizon with the bifurcation surface on the background spacetime when the number of the null vector $k^a$ is less than or equal to the null vector $l^a$ \cite{Wang:2022stg, Sang:2021rla}. In other words, the quantity $X_{a_1 \cdots a_k} z_{1}^{a_1} \cdots z_{k}^{a_k}$ is a quantity on the background spacetime when the number of $k^a$ is less than or equal to the number of $l^a$, and the quantity is a quantity under the first-order approximation when the number of $k^a$ is larger than the number of $l^a$. This result can be regarded as a criterion to judge whether the quantities in $H_{uu}$ are contributed under the zero-order approximation or only under the first-order approximation. For simplicity, the quantity under the zeroth-order approximation (or on the background spacetime) is called the ``background quantity'', and the quantity under the linear order approximation is called the ``first-order quantity'' directly. Moreover, in the following calculations, we will use the symbols $( \quad )_n$ or $[ \quad ]_n\,, \, (n = 0, 1)$ to label the quantities in $H_{uu}$ during the calculation process, where $n = 0$ represents the background quantity and $n = 1$ represents the first-order quantity.

After contracting two null vectors $k^a$ with the first component on the left-hand side of the equation of motion, the first identity in Eq. (\ref{specexprefourpartofeom}) can be expressed as
\begin{equation}\label{firsthkk}
	H_{uu}^{1} = k^a k^b (E_R)_{acde} R_b^{\ cde} + 2 k^a k^b \nabla^c \nabla^d \left(E_R \right)_{acbd}\,.
\end{equation}
Expanding the repeated indexes by using the definition of the induced metric on the cross-section of the event horizon, the specific expression of the first term of Eq. (\ref{firsthkk}) under the first-order approximation of the perturbation process is further given as
\begin{equation}\label{ffhkk}
	\begin{split}
		& k^a k^b (E_R)_{acde} R_b^{\ cde} \\
		= & \left[(E_R)_{u i j k} \right]_1 \left(R_{u}^{\ i j k} \right)_1 - 2 \left[(E_R)_{u i v j} \right]_0 \left(R^{v i v j} \right)_1 - 2 \left[(E_R)_{u i u j} \right]_1 \left(R^{v i u j} \right)_0 \\
		& + 2 \left[(E_R)_{u i v u} \right]_1 \left(R_{u v u}^{\ \ \ i} \right)_1\\
		\simeq & - 2 (E_R)_{u i v j} R^{v i v j} - 2 (E_R)_{i u j u}  R^{v i u j} \,.
	\end{split}
\end{equation}
From the Lagrangian of the gravitational theory, the quantity $(E_R)_{i u j u}$ under the first-order approximation can be further calculated as 
\begin{equation}\label{ekkpab}
	\begin{split}
		(E_R)_{i u j u} \simeq & 4 \frac{\partial^2 \mathcal{L}}{\partial R^{v k v l} \partial R^{u i u j}} R^{k v l v} + \frac{\partial^2 \mathcal{L}}{\partial \left(\nabla_u \nabla_u \phi \right) \partial R^{u i u j}} \nabla_u \nabla_u \phi \simeq \mathcal{L}_k \mathcal{P}_{i j}\,,
	\end{split}
\end{equation}
where
\begin{equation}
	\mathcal{P}_{ij} = - 4 \frac{\partial^2 \mathcal{L}}{\partial R^{v k v l} \partial R^{u i u j}} B^{k l} + \frac{\partial^2 \mathcal{L}}{\partial \left(\nabla_u \nabla_u \phi \right) \partial R^{u i u j}} \mathcal{L}_k \phi\,.
\end{equation}
Using the identities in Eq. (\ref{ekkpab}) and Eq. (\ref{evolutionc}), the result of Eq. (\ref{ffhkk}) under the linear order approximation can be further simplified as  
\begin{equation}\label{rrefirsthkk}
	\begin{split}
		k^a k^b (E_R)_{acde} R_b^{\ cde} \simeq 2 \mathcal{L}_k \left(\mathcal{P}_{i j} \mathcal{L}_k C^{i j} \right) - 2 (E_R)_{u i v j} R^{v i v j}\,.
	\end{split}
\end{equation}
Utilizing Leibniz's law, the second term of Eq. (\ref{firsthkk}) can be expanded as
\begin{equation}\label{secondhkk}
	\begin{split}
		2 k^a k^b \nabla^c \nabla^d \left(E_R \right)_{acbd} = 2 k^a \nabla^c \left(k^b \nabla^d \left(E_R \right)_{acbd} \right) - 2 \left( k^a \nabla^c k^b \right) \nabla^d \left(E_R \right)_{acbd}\,.
	\end{split}
\end{equation}
Since we discuss the linearized second law of thermodynamics in general diffeomorphism invariant high-order curvature gravitational theory with the scalar and the electromagnetic fields, the specific expression of the quantity $\left(E_R \right)_{abcd}$ cannot be obtained directly. It means that we cannot obviously extract the expression of the first term on the right-hand side of Eq. (\ref{secondhkk}) under the first-order approximation of the perturbation process. Therefore, a significant identity should be involved first to effectively obtain the expression of this term under the linear order approximation. From any two-form tensor $X^{ab}$, we can demonstrate that the identity can be given as
\begin{equation}\label{generalrelation}
	\int_s \tilde{\epsilon} k_b \nabla_a X^{ab} = \frac{1}{2} \mathcal{L}_k \int \epsilon_{a b a_1 \cdots a_{n}} X^{ab}\,.
\end{equation}
Using this identity twice and Leibniz's law, while according to the density of the Wald entropy in Eq. (\ref{entropydensity}), the integral form of the second term of Eq. (\ref{firsthkk}) on the cross-section of the event horizon can be further written as
\begin{equation}\label{firstsecondhkk}
	\begin{split}
		& 2 \int_s \tilde{\epsilon} k^a k^b \nabla^c \nabla^d \left(E_R \right)_{acbd} \\
		= & - \mathcal{L}_k^2 \int_s \tilde{\epsilon} \rho_W  - 2 \int_s \tilde{\epsilon} \left( k^a \nabla^c k^b \right) \nabla^d \left(E_R \right)_{acbd} + 2 \mathcal{L}_k \int_s \tilde{\epsilon} \left[ \left(k^b \nabla^d l^a \right) k^c \left(E_R \right)_{cabd} \right] \\
		& + 2 \mathcal{L}_k \int_s \tilde{\epsilon} \left[ \left(k^b \nabla^d k^c \right) l^a \left(E_R \right)_{cabd} \right]\,.
	\end{split}
\end{equation}
Before calculating the above expression under the linear order approximation, three practical identities are introduced to simplify the calculation process, and these identities have been demonstrated in our previous research work \cite{Wang:2021zyt}. The three identities in the background spacetime can be expressed as
\begin{equation}\label{usefulthreeident}
	\nabla_u l^a = 0\,, \quad \nabla_i k_a = 0\,, \quad \nabla_a k_i = 0\,.
\end{equation}
Meanwhile, the three identities also indicate that the three quantities on the left-hand side of each identity are all first-order quantities.

For the integrand of the second integral in Eq. (\ref{firstsecondhkk}), using the three identities in Eq. (\ref{usefulthreeident}), we can expand the repeated indexes and calculate the specific expression under the first-order approximation as 
\begin{equation}\label{seintsecondhkk}
	\begin{split}
		& \left( k^a \nabla^c k^b \right) \nabla^d \left(E_R \right)_{acbd} \\
		= & - \left[\nabla_u \left(E_R \right)_{u i v d}  \right]_1 \left(\nabla^{i} k^d \right)_1 - \left[\nabla_v \left(E_R \right)_{u i u d} \right]_1 \left(\nabla^{i} k^d \right)_1 - \left(l_a \nabla^{i} k^a \right)_1 \left[\nabla^{j} \left(E_R \right)_{u i j u} \right]_1\\
		& + \left(B^{i j} \right)_1 \left[\nabla^{k} \left(E_R \right)_{u i k j} \right]_1\\
		\simeq & 0
	\end{split}
\end{equation}
The integrand in the third integral of Eq. (\ref{firstsecondhkk}) can be further calculated as
\begin{equation}\label{thirdintsecondhkk}
	\begin{split}
		& \left(k^b \nabla^d l^a \right) k^c \left(E_R \right)_{cabd} \\
		= & - \left[\left(E_R \right)_{u i u v} \right]_1 \left(\nabla_u l^{i} \right)_1 + \left[\left(E_R \right)_{u i u j} \right]_1 \left(C^{i j} \right)_0 + \left[\left(E_R \right)_{u v u i} \right]_1 \left(l^a \nabla^{i} k_a \right)_1 - \frac{1}{2} \rho_{\text{W}} k^a \nabla_u l_a\\
		\simeq & \left(E_R \right)_{u i u j} C^{i j}\,,
	\end{split}
\end{equation}
under the linear order approximation of the matter fields perturbation. The last term in the second line of Eq. (\ref{thirdintsecondhkk}) is equal to zero directly according to the geodesic equation $k^b \nabla_b k^a = 0$. Since the integrand in the fourth term of Eq. (\ref{firstsecondhkk}) under the first-order approximation of the perturbation process is 
\begin{equation}\label{fourintsecondhkk}
	\begin{split}
		\left(k^b \nabla^d k^c \right) l^a \left(E_R \right)_{cabd} = & - \left[\left(E_R \right)_{v i u j} \right]_0 \left(B^{i j} \right)_1 - \left[\left(E_R \right)_{u v u i} \right]_1 \left(l^c \nabla^{i} k_c \right)_1\\
		\simeq & - \left(E_R \right)_{v i u j} B^{i j}\,,
	\end{split}
\end{equation}
the fourth term in the result of Eq. (\ref{firstsecondhkk}) can be finally written as
\begin{equation}\label{fourthintsecondhkk}
	\begin{split}
		2 \mathcal{L}_k \int_s \tilde{\epsilon} \left[ \left(k^b \nabla^d k^c \right) l^a \left(E_R \right)_{cabd} \right] \simeq 2 \int_s \tilde{\epsilon} R^{i v j v} \left(E_R \right)_{v i u j}\,,
	\end{split}
\end{equation}
where the second identity in Eq. (\ref{bandevlob}) is used in the last step. 

In conclusion, utilizing the results in Eqs. (\ref{rrefirsthkk}), (\ref{seintsecondhkk}), (\ref{thirdintsecondhkk}), and (\ref{fourthintsecondhkk}), the expression of the integral of $H_{uu}^{1}$ on the cross-section under the linear order approximation of the perturbation process can be finally obtained as
\begin{equation}\label{huu1final}
	\begin{split}
		\int_s \tilde{\epsilon} H_{uu}^{1} \simeq & - \mathcal{L}_k^2 \int_s \tilde{\epsilon} \rho_W + 2 \mathcal{L}_k \int_s \tilde{\epsilon} \left(E_R \right)_{u i u j} C^{i j} + 2 \mathcal{L}_k \int_s \tilde{\epsilon} \left(\mathcal{P}_{i j} \mathcal{L}_k C^{i j} \right) \\
		& + 2 \int_s \tilde{\epsilon} R^{i v j v} \left(E_R \right)_{v i u j} - 2 \int_s \tilde{\epsilon} (E_R)_{u i v j} R^{v i v j}\\
		= & - \mathcal{L}_k^2 \int_s \tilde{\epsilon} \rho_W + 2 \mathcal{L}_k \int_s \tilde{\epsilon} \left(\mathcal{L}_k \mathcal{P}_{i j} \right) C^{i j} + 2 \mathcal{L}_k \int_s \tilde{\epsilon} \left(\mathcal{P}_{i j} \mathcal{L}_k C^{i j} \right) \\
		= & - \mathcal{L}_k^2 \int_s \tilde{\epsilon} \left(\rho_W - 2 \mathcal{P}_{i j} C^{i j} \right)\,,
	\end{split}
\end{equation}
where we have used Eq. (\ref{ekkpab}) in the second step.

For the second identity in Eq. (\ref{specexprefourpartofeom}), after contracting two null vectors $k^a$ with the expression of $H_{ab}^{2}$, the expression of $H_{uu}^2$ under the linear order approximation can be directly calculated as 
\begin{equation}\label{huu2final}
	\begin{split}
		H_{uu}^2 = \frac{1}{2} k^a k^b \left(E_1 \right)_a \nabla_b \phi = \frac{1}{2} \left[\left(E_1 \right)_u \right]_1 \left(\mathcal{L}_k \phi \right)_1 \simeq 0\,.
	\end{split}
\end{equation}
Since $H_{uu}^{2}$ is vanish under the first-order approximation of the perturbation, the integral on the cross-section of the event horizon does not contribute to the final result.

Contracting two null vectors $k^a$ with the third identity in Eq. (\ref{specexprefourpartofeom}), the expression of $H_{uu}^3$ can be given as 
\begin{equation}
	H_{uu}^3 = - k^a k^b \nabla^c \left(E_2 \right)_{cb} \nabla_a \phi + \frac{1}{2} k^a k^b \nabla^c \left(E_2 \right)_{ab} \nabla_c \phi + \frac{1}{2} k^a k^b \left(E_2 \right)_{ab} \nabla_c \nabla^c \phi\,.
\end{equation}
After expanding the repeated indexes using the definition of the induced metric on the cross-section as well, the expression of $H_{uu}^3$ under the first-order approximation can be further expressed as 
\begin{equation}\label{hkk3}
	\begin{split}
		H_{uu}^3 = & - \left[ \left(E_2 \right)_{u u} \right]_1 \left(\nabla_v \nabla_u \phi \right)_0 + \left(\mathcal{L}_k \phi \right)_1 \left[\nabla_u \left(E_2 \right)_{uv} \right]_1 + \frac{1}{2} \left(D^{i} \phi \right)_0 \left[\nabla_{i} \left(E_2 \right)_{uu} \right]_1 \\
		& + \frac{1}{2} \left(\mathcal{L}_k \phi \right)_1 \left[\nabla_v \left(E_2 \right)_{u u} \right]_1 - \left(\mathcal{L}_k \phi \right)_1 \left[\nabla^{i} \left(E_2 \right)_{u i} \right]_1 - \frac{1}{2} \left[\nabla_u \left(E_2 \right)_{u u} \right]_1 \left(\nabla_v \phi \right)_0\\
		& + \frac{1}{2} \left[\left(E_2 \right)_{u u} \right]_1 \left(D^{i} D_{i} \phi \right)_0\\
		\simeq & \frac{1}{2} \left(D^{i} \phi \right) \nabla_{i} \left(E_2 \right)_{uu} + \frac{1}{2} \left(E_2 \right)_{u u} \left(D^{i} D_{i} \phi \right) - \left(E_2 \right)_{u u} \left(\nabla_v \nabla_u \phi \right) \\
		& - \frac{1}{2} \left(\nabla_v \phi \right) \nabla_u \left(E_2 \right)_{u u}\,.
	\end{split}
\end{equation}
The first term of the result in Eq. (\ref{hkk3}) is further calculated as 
\begin{equation}\label{firstfohkk3}
	\begin{split}
	& \frac{1}{2} \left(D^{i} \phi  \right) \nabla_{i} \left(E_2 \right)_{u u}\\
	= & \frac{1}{2} \left(D^{i} \phi \right)_0 D_{i} \left[\left(E_2 \right)_{u u} \right]_1 - \frac{1}{2} \left(D_{i} \phi \right)_0 \left[\left(E_2 \right)_{u j} \right]_1 \left(B^{i j} \right)_1 + \frac{1}{2} \left(D^{i} \phi \right)_0 \left[\left(E_2 \right)_{u u} \right]_1 \left(l^d \nabla_{i} k_d \right)_1 \\
	& - \frac{1}{2} \left(D_{i} \phi \right)_0 \left[\left(E_2 \right)_{j u} \right]_1 \left(B^{i j} \right)_1 + \frac{1}{2} \left(D^{i} \phi \right)_0 \left[\left(E_2 \right)_{u u} \right]_1 \left(l^d \nabla_{i} k_d \right)_1\\
	\simeq & \frac{1}{2} \left(D^{i} \phi \right) D_{i} \left[\left(E_2 \right)_{u u} \right]
	\end{split}
\end{equation}
under the first-order approximation of the matter field perturbation. Combining the result of Eq. (\ref{firstfohkk3}) with the second term of the result in Eq. (\ref{hkk3}), we have
\begin{equation}\label{firsttwotermsfohkk3}
	\begin{split}
		\frac{1}{2} \left(D^{i} \phi \right) D_{i} \left[\left(E_2 \right)_{u u} \right] + \frac{1}{2} \left(E_2 \right)_{u u} \left(D^{i} D_{i} \phi \right) = \frac{1}{2} D_{i} \left[\left(D^{i} \phi \right) \left(E_2 \right)_{u u} \right]\,.
	\end{split}
\end{equation}
When the topology of the event horizon of black holes in the general high-order curvature gravitational theory with the scalar and the electromagnetic fields is assumed to be compact, the spatial total derivative term in the integrand, namely the spatial boundary term of the integral, does not contribute to the final result. Therefore, the result of Eq. (\ref{firsttwotermsfohkk3}) can be neglected directly. According to the Lagrangian of the gravitational theory, the expression of $\left(E_2 \right)_{u u}$ under the first-order approximation can be further given as
\begin{equation}\label{E2uu}
	\begin{split}
		\left(E_2 \right)_{u u} \simeq & \frac{\partial^2 \mathcal{L}}{\partial \left(\nabla_u \nabla_u \phi \right) \partial \left(\nabla_v \nabla_v \phi \right) } \nabla_u \nabla_u \phi + 4 \frac{\partial^2 \mathcal{L}}{\partial R^{v i v j} \partial \left(\nabla_v \nabla_v \phi \right)} R^{i v j v} \simeq \mathcal{L}_k \mathcal{N}\,,
	\end{split}
\end{equation}
where
\begin{equation}
	\mathcal{N} = \frac{\partial^2 \mathcal{L}}{\partial \left(\nabla_u \nabla_u \phi\right) \partial \left(\nabla_v \nabla_v \phi \right)} \mathcal{L}_k \phi - 4 \frac{\partial^2 \mathcal{L}}{\partial R^{v i v j} \partial \left(\nabla_v \nabla_v \phi \right)} B^{i j}\,.
\end{equation}
Using the identity in Eq. (\ref{E2uu}), the third and fourth terms in the result of Eq. (\ref{hkk3}) under the linear order approximation can be further simplified as
\begin{equation}
	\begin{split}
		- \left[\left(E_2 \right)_{u u} \right] \left(\nabla_v \nabla_u \phi \right) - \frac{1}{2} \left[\nabla_u \left(E_2 \right)_{u u} \right] \left(\nabla_v \phi \right) \simeq - \mathcal{L}_k^2 \left[\frac{1}{2} \mathcal{N} \left(\nabla_v \phi \right) \right]\,.
	\end{split}
\end{equation}
Therefore, the expression of the integral of $H_{uu}^3$ on the cross-section under the first-order approximation is given as
\begin{equation}\label{huu3final}
	\int_s \tilde{\epsilon} H_{uu}^3 \simeq - \mathcal{L}_k^2 \int_s \tilde{\epsilon} \left[\frac{1}{2} \mathcal{N} \left(\nabla_v \phi \right) \right]\,.
\end{equation}

Finally, after contracting two null vectors $k^a$ with the fourth identity in Eq. (\ref{specexprefourpartofeom}) and expanding the repeated index by the induced metric, the expression of $H_{kk}^4$ under the first-order approximation of the perturbation process is directly given as 
\begin{equation}\label{huu4final}
		H_{uu}^{4} = k^a k^b \left(E_F \right)_{ac} F_{b}^{\ c} = \left[\left(E_F \right)_{u i} \right]_1 \left(F_{u}^{\ i} \right)_1 \simeq 0\,.
\end{equation}
The result shows that the electromagnetic field part does not contribute to the general expression of the entropy of black holes satisfying the linearized second law in the gravitational theory under the first-order approximation.

Therefore, combining the results in Eqs. (\ref{huu1final}), (\ref{huu2final}), (\ref{huu3final}), and (\ref{huu4final}), and supplementing the coefficient $1/4$ on two sides of the identity, while utilizing the definition of the Wald entropy in Eq. (\ref{definewaldentropy}), the equation of motion of the gravitational part in Eq. (\ref{equationofmotion}) and the null energy condition in Eq. (\ref{entropydensity}), we have
\begin{equation}
	\begin{split}
		\mathcal{L}_k^2 \left(S_{\text{W}} + S_{\text{ct}} \right) \simeq - \frac{1}{4} \int_s \tilde{\epsilon} H_{u u} = - 2 \pi \int_s \tilde{\epsilon} T_{uu} \le 0\,,
	\end{split}
\end{equation}
where 
\begin{equation}\label{expresssct}
	S_{\text{ct}} = \frac{1}{4} \int_s \tilde{\epsilon} \left[- 2 \mathcal{P}_{i j} C^{i j} + \frac{1}{2} \mathcal{N} \left(\nabla_v \phi \right) \right]\,.
\end{equation}
The result shows that the general expression of the entropy of black holes obeying the linearized second law in any diffeomorphism invariant high-order curvature gravitational theory with the scalar and the electromagnetic fields during the matter fields perturbation process can be written as the Wald entropy with two correction terms eventually. Moreover, the two correction terms include not only the contribution from the non-minimal coupling interactions in gravitational theory but also the contribution from the minimal coupling interaction between gravity and the scalar fields. And the contribution of the minimal coupling interaction is contained in the expression of $\mathcal{N}$ in the second term of Eq. (\ref{expresssct}). According to the result in previous literature, the general expression of the entropy of black holes that always satisfies the linearized second law in an arbitrary high-order curvature gravitational theory can be expressed as the form of the Wald entropy with correction terms. These correction terms only come from the contribution of the non-minimal coupling self-interaction of gravity. Besides, for the gravitational theory with matter fields used to study the linearized second law in our previous research works, the entropy of black holes obeying the linearized second law can be expressed as the Wald entropy or the Wald entropy with some correction terms. These correction terms also only come from the non-minimal coupling interaction terms between gravity and matter fields, while the minimal coupling interactions do not influence the specific expression of the entropy. However, when considering the general diffeomorphism invariant high-order curvature gravity with the scalar and the electromagnetic fields, the minimal coupling interaction between gravity and the scalar fields will contribute to the correction terms of the Wald entropy. This result overturns our previous understanding of the expression of the entropy of black holes satisfying the linearized second law of thermodynamics in an arbitrary diffeomorphism invariant gravity.

\section{Discussions and Conclusions}\label{sec4}
Since an appropriate scheme to quantize gravity has not been established until now, the low energy efficient theory corresponding to the gravitational theory should be constructed first to approximatively study the properties of the gravitational theory at the quantum level. When investigating the quantum gravitational theory, we are mainly interested in various interactions in the gravitational theory. These interactions can be classified into two categories. One is the self-interaction of gravity, and the other is the interactions between gravity and matter fields. When the gravitational theory involves the two types of interactions, the quantum correction terms, which correspond to the two types of interactions, will be added to the expression of the Lagrangian of the low-energy efficient gravitational theory. For any diffeomorphism invariant gravitational theory with the two types of interaction, it has been demonstrated that the first law of black holes can be generally constructed, and the entropy in the expression of the first law is called the Wald entropy rather than the Bekenstein-Hawking entropy. It indicates that the Wald entropy is commonly suitable for the first law of black hole thermodynamics in any diffeomorphism invariant gravitational theory.

However, whether the Wald entropy is commonly suitable for the second law of black hole thermodynamics should be further investigated. According to the definition of the Wald entropy, when the non-minimal coupling interaction terms describe the two types of interactions added in the Lagrangian of the theory of gravity, these terms will sufficiently affect the expression of the Wald entropy. Therefore, to generally investigate whether the Wald entropy satisfies the second law, we should first examine whether the Wald entropy in the gravitational theory with specific non-minimal coupling interaction obeys the second law. For the case that the Lagrangian of the gravitational theory only contains the high-order curvature terms, namely the Lagrangian containing the non-minimal coupling self-interaction of gravity, this situation has been investigated generally, and the result shows that the Wald entropy no longer obeys the requirement of the second law under the first-order approximation of the matter field perturbation. The general expression of the entropy of black holes obeying the linearized second law in this gravitational theory can be expressed as the Wald entropy with some correction terms. Although the linearized second law in any high-order curvature gravity has been investigated adequately, the second law in the gravitational theory with matter fields has not been studied enough. Since only the non-minimal coupling interactions of gravity with matter fields can affect the expression of the Wald entropy, we only focus on the linearized second law of black hole thermodynamics in the gravitational theory containing the non-minimal coupling interaction between gravity and matter fields. The linearized second law of black holes in Horndeski gravity, general quadratic corrected Einstein-Maxwell gravity, and general second-order scalar-tensor gravity have been investigated in our previous works. The results show that the Wald entropy also no longer commonly satisfies the requirement of the linearized second law, and the entropy of black holes that obeys the linearized second law can be expressed as the Wald entropy with some correction terms as well. Based on these previous works, we would like to investigate the linearized second law of black holes in general diffeomorphism invariant high-order curvature gravitational theory with arbitrary matter fields and to obtain the entropy of black holes that generally satisfies the linearized second law during the perturbation process.

Considering a general diffeomorphism invariant high-order curvature gravitational theory with the scalar fields and the Maxwell fields, starting with the definition of the Wald entropy again, we would like to derive the general expression of the entropy of black holes in the theory of gravity that always satisfies the linearized second law of black hole thermodynamics. A quasistationary accreting process should be introduced first, which states that the additional matter fields minimally coupling with gravity outside black holes pass through the event horizon and fall into black holes. It means that the matter fields can perturb the spacetime configuration of black holes during the accreting process. Furthermore, two assumptions are introduced to investigate the linearized second law. The first is that the additional matter fields obey the null energy condition, and the second is the stability assumption. Besides, to ensure that all physical quantities are smooth and finite on the whole Killing horizon, the third assumption that a regular bifurcation surface exists in the background spacetime is also introduced. From the Raychaudhuri equation, according to the regularity of the bifurcation surface and the null energy condition of additional matter fields, the general expression of the entropy of black holes obeying linearized second law in any high-order gravitational theory with the scalar and the electromagnetic fields during the perturbation process is obtained. This entropy can also be written as the Wald entropy with two correction terms. This result overturns our previous understanding of the entropy of black holes that obeys the linearized second law because the contribution of the minimal coupling interaction between gravity and the scalar field is contained in one of the two correction terms. Meanwhile, this result leads us to a new understanding of the entropy of black holes satisfying the second law in any diffeomorphism invariant gravitational theory with arbitrary matter fields. The entropy of black holes obeying the linearized second law in any gravitational theory with matter fields contains not only the contribution from the non-minimal coupling interaction in the gravitational theory but also the contribution from the minimal coupling interaction between gravity with matter fields.

\section*{Acknowledgement}
X.-Y. W. is supported by the National Natural Science Foundation of China  with Grant No. 12105015 and the Talents Introduction Foundation of Beijing Normal University with Grant No. 111032109. J. J. is supported by the GuangDong Basic and Applied Basic Research Foundation with Grant No. 217200003 and the Talents Introduction Foundation of Beijing Normal University with Grant No. 310432102.

\end{document}